\documentclass[twoside,twocolumn,9pt]{article}
\usepackage{amsmath}
\usepackage{amssymb}
\usepackage{extsizes}
\usepackage[super,sort&compress,comma]{natbib} 
\usepackage[version=3]{mhchem}
\usepackage[left=1.5cm, right=1.5cm, top=1.785cm, bottom=2.0cm]{geometry}
\usepackage{balance}
\usepackage{mathptmx}
\usepackage{sectsty}
\usepackage{graphicx} 
\usepackage{lastpage}
\usepackage[format=plain,justification=justified,singlelinecheck=false,font={stretch=1.125,small,sf},labelfont=bf,labelsep=space]{caption}
\usepackage{float}
\usepackage{fancyhdr}
\usepackage{fnpos}
\usepackage[english]{babel}
\addto{\captionsenglish}{%
  
}
\usepackage{array}
\usepackage{droidsans}
\usepackage{charter}
\usepackage[T1]{fontenc}
\usepackage[usenames,dvipsnames]{xcolor}
\usepackage{setspace}
\usepackage[compact]{titlesec}
\usepackage{hyperref}
\definecolor{cream}{RGB}{222,217,201}

\begin{document}

\pagestyle{fancy}
\thispagestyle{plain}
\fancypagestyle{plain}{
%%%HEADER%%%
\renewcommand{\headrulewidth}{0pt}
}
%%%END OF HEADER%%%

%%%PAGE SETUP - Please do not change any commands within this section%%%
\makeFNbottom
\makeatletter
\renewcommand\LARGE{\@setfontsize\LARGE{15pt}{17}}
\renewcommand\Large{\@setfontsize\Large{12pt}{14}}
\renewcommand\large{\@setfontsize\large{10pt}{12}}
\renewcommand\footnotesize{\@setfontsize\footnotesize{7pt}{10}}
\makeatother

\renewcommand{\thefootnote}{\fnsymbol{footnote}}
\renewcommand\footnoterule{\vspace*{1pt}% 
\color{cream}\hrule width 3.5in height 0.4pt \color{black}\vspace*{5pt}} 
\setcounter{secnumdepth}{5}

\makeatletter 
\renewcommand\@biblabel[1]{#1}            
\renewcommand\@makefntext[1]% 
{\noindent\makebox[0pt][r]{\@thefnmark\,}#1}
\makeatother 
\renewcommand{\figurename}{\small{Fig.}~}
\sectionfont{\sffamily\Large}
\subsectionfont{\normalsize}
\subsubsectionfont{\bf}
\setstretch{1.125} %In particular, please do not alter this line.
\setlength{\skip\footins}{0.8cm}
\setlength{\footnotesep}{0.25cm}
\setlength{\jot}{10pt}
\titlespacing*{\section}{0pt}{4pt}{4pt}
\titlespacing*{\subsection}{0pt}{15pt}{1pt}
%%%END OF PAGE SETUP%%%

%%%FOOTER%%%
\fancyfoot{}
\fancyfoot[LO,RE]{\vspace{-7.1pt}\includegraphics[height=9pt]{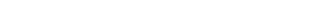}}
\fancyfoot[CO]{\vspace{-7.1pt}\hspace{13.2cm}\includegraphics{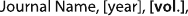}}
\fancyfoot[CE]{\vspace{-7.2pt}\hspace{-14.2cm}\includegraphics{head_foot/RF}}
\fancyfoot[RO]{\footnotesize{\sffamily{1--\pageref{LastPage} ~\textbar  \hspace{2pt}\thepage}}}
\fancyfoot[LE]{\footnotesize{\sffamily{\thepage~\textbar\hspace{3.45cm} 1--\pageref{LastPage}}}}
\fancyhead{}
\renewcommand{\headrulewidth}{0pt} 
\renewcommand{\footrulewidth}{0pt}
\setlength{\arrayrulewidth}{1pt}
\setlength{\columnsep}{6.5mm}
\setlength\bibsep{1pt}
%%%END OF FOOTER%%%

%%%FIGURE SETUP - please do not change any commands within this section%%%
\makeatletter 
\newlength{\figrulesep} 
\setlength{\figrulesep}{0.5\textfloatsep} 

\newcommand{\topfigrule}{\vspace*{-1pt}% 
\noindent{\color{cream}\rule[-\figrulesep]{\columnwidth}{1.5pt}} }

\newcommand{\botfigrule}{\vspace*{-2pt}% 
\noindent{\color{cream}\rule[\figrulesep]{\columnwidth}{1.5pt}} }

\newcommand{\dblfigrule}{\vspace*{-1pt}% 
\noindent{\color{cream}\rule[-\figrulesep]{\textwidth}{1.5pt}} }

\makeatother
%%%END OF FIGURE SETUP%%%

%%%TITLE, AUTHORS AND ABSTRACT%%%
\twocolumn[
  \begin{@twocolumnfalse}
{\includegraphics[height=30pt]{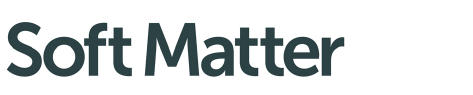}\hfill\raisebox{0pt}[0pt][0pt]{\includegraphics[height=55pt]{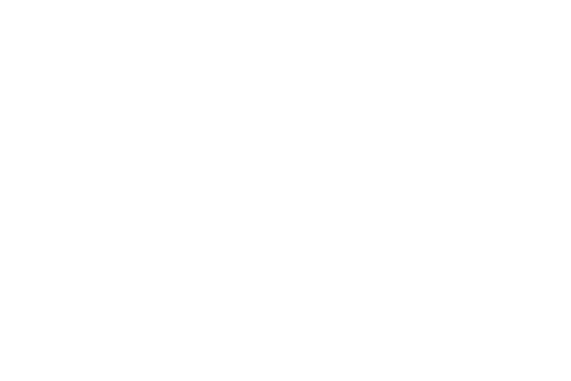}}\\[1ex]
\includegraphics[width=18.5cm]{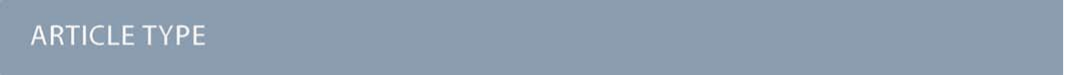}}\par
\vspace{1em}
\sffamily
\begin{tabular}{m{4.5cm} p{13.5cm} }

\includegraphics{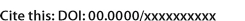} & \noindent\LARGE{\textbf{Hysteretic wavelength selection in isometric, unsupported radial wrinkling
%$^\dag$
}} \\%Article title goes here instead of the text "This is the title"
\vspace{0.3cm} & \vspace{0.3cm} \\

 & \noindent\large{Anshuman S. Pal$^{\ast}$\textit{$^{a}$} } \\%Author names go here instead of "Full name", etc.

\includegraphics{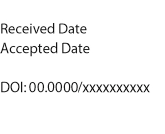} & \noindent\normalsize{
In [Pal \textit{et al.}, arXiv:2206.03552], the authors discuss how an unsupported flat annulus contracted at its inner boundary by fraction $\Delta$, buckles into a radial wrinkling pattern that is asymptotically isometric and tension-free. What selects the wavelength in such a pure-bending configuration, in the absence of any competing sources of work? In this paper, with the support of numerical simulations, we argue that competition between stretching and bending energies at local, mesoscopic scales leads to the selection of a wavelength scale $\lambda^*$ sensitive to both the width $w$ and thickness $t$ of the sheet: $\lambda^* \sim w^{2/3} t^{1/3} \Delta^{-1/6}$. This scale $\lambda^*$ corresponds to a kinetic arrest criterion for wrinkle coarsening starting from any  finer wavelength $\lambda \lesssim \lambda^*$.
%, which can be interpreted in terms of both size and energetic barriers to further coarsening. 
However, the sheet can support coarser wavelengths: $\lambda \gtrsim \lambda^*$, since there is no penalty to their existence. Since this wavelength selection mechanism depends on the initial value of $\lambda$, it is path-dependent or hysteretic.} \\%The abstrast goes here instead of the text "The abstract should be..."

\end{tabular}

 \end{@twocolumnfalse} \vspace{0.6cm}

  ]
%%%END OF TITLE, AUTHORS AND ABSTRACT%%%

%%%FONT SETUP - please do not change any commands within this section
\renewcommand*\rmdefault{bch}\normalfont\upshape
\rmfamily
\section*{}
\vspace{-1cm}

%%%FOOTNOTES%%%

\footnotetext{\textit{$^{a}$~Department of Physics, University of Chicago, Chicago, IL 60637, USA.  E-mail: anshuman@uchicago.edu}}
\footnotetext{\textit{$^\ast$}~Corresponding author}
%\footnotetext{\textit{$^{b}$~Address, Address, Town, Country. }}

%Please use \dag to cite the ESI in the main text of the article.
%If you article does not have ESI please remove the the \dag symbol from the title and the footnotetext below.
%\footnotetext{\dag~Electronic Supplementary Information (ESI) available: [details of any supplementary information available should be included here]. See DOI: 10.1039/cXsm00000x/}
%additional addresses can be cited as above using the lower-case letters, c, d, e... If all authors are from the same address, no letter is required

%\footnotetext{\ddag~Additional footnotes to the title and authors can be included \textit{e.g.}\ `Present address:' or `These authors contributed equally to this work' as above using the symbols: \ddag, \textsection, and \P. Please place the appropriate symbol next to the author's name and include a \texttt{\textbackslash footnotetext} entry in the the correct place in the list.}

%%%END OF FOOTNOTES%%%
%%%%%%%%%%%%%%%%%%%%%%%%%%%%%%%%%%%%%%%%%%
%%%MAIN TEXT%%%%
Thin elastic sheets under confinement buckle and bend to form a multitude of shapes. Perhaps the most ubiquitous of these forms of elastic pattern formation is \textit{wrinkling}, viz.~where the excess material length gets collected in multiple undulations. A central concern in the study of these structures is the determination of the wavelength $\lambda$ of these undulations -- which constitutes an \textit{emergent} intermediate length scale, much smaller than the system size but much larger than the sheet thickness. Usually, this wavelength is selected \cite{Paulsen2019} through competition between the sheet's bending stiffness and some external source of stiffness or deformation work, like a substrate \cite{Bowden1998}, a tension field \cite{Cerda2003, Davidovitch2011}, inertia \cite{Box2019}, or even extrinsic curvature \cite{Paulsen2016}. While the bending stiffness \textit{promotes} larger $\lambda$ in order to minimise wrinkle curvature, the substrate or other source of work tends to \textit{penalise} large $\lambda$ since it leads to large amplitude deformation. But how is wavelength determined in an unsupported sheet, i.e.~in the absence of such external sources of stiffness?
\par
\textit{A priori}, an unconstrained sheet under compression should spontaneously choose the maximum wavelength possible -- at the scale of the system size -- in order to minimise bending energy. Thus, the minimum ingredient for generating an intermediate wavelength is the presence of some external \textit{constraint}. Besides a substrate, another possible source of constraints is clamping at the boundary. References \cite{Davidovitch2009, Huang2010, Schroll2011, Vandeparre2011} study such systems where relatively coarse wrinkling in the bulk of the sheet gradually becomes refined in the proximity of a clamped or pinned boundary, in order to minimise the wrinkling amplitude. Of particular interest to us are references \cite{Vandeparre2011, Pomeau1997}, in particular  Vandeparre et al.~\cite{Vandeparre2011}, which consider wrinkling in an \textit{unsupported} rectangular sheet contracted at one boundary. Here, the wrinkle wavelength in the bulk is determined by the wavelength fixed (i.e.~clamped) at the boundary, coarsening outward through a `wrinkle hierarchy` (see Figs.~\ref{fig:fig2}b and \ref{fig:fig4}a). But what if the sheet is \textit{unsupported and also unclamped}? 
%%%%%%%%%%%%%%%%%%%%
%%%%%%%%%%%%%%%%%%%%
\begin{figure*}[htb]
\centering
\includegraphics[width=0.8\textwidth]{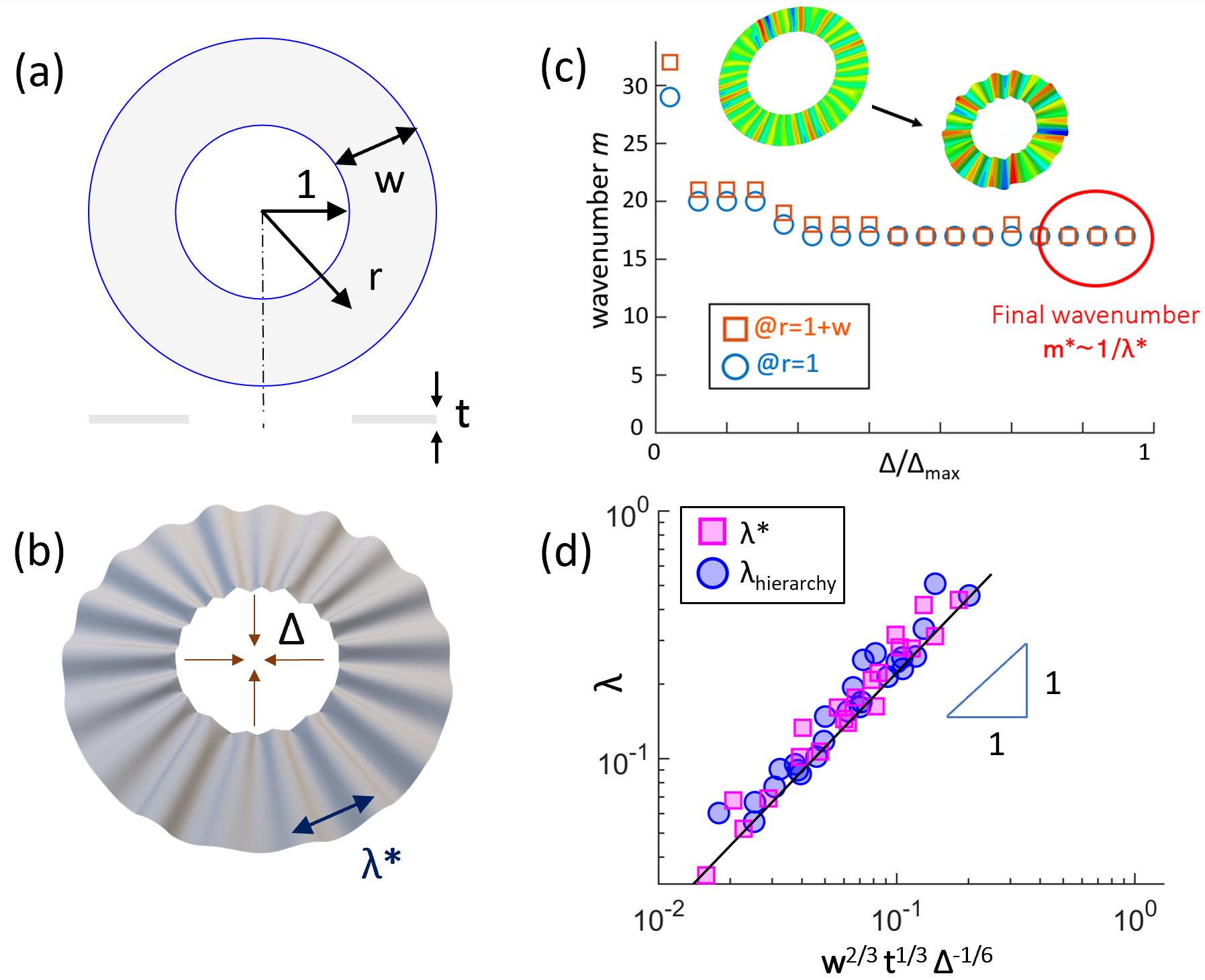}
\caption[Geometry of the Inner Lamé radially wrinkled system]
{Geometry of the Inner Lamé radially wrinkled system, and selected wavelengths in its numerical implementation.
(a) A schematic diagram of the flat annulus, showing its geometric parameters: radial distance $r$, width $w$ and (in cross-section) thickness $t$. We take the inner radius as unity.
(b) An example of a deformed configuration, showing the displacement $\Delta$ and the wrinkle wavelength $\lambda^*$, measured at the outer boundary, which we wish to predict.
(c) shows representative data for the evolution of the wavenumber (i.e.~the number of wrinkles) against normalised contraction $\Delta/\Delta_{\rm max}$, measured at both inner ($r=1$) and outer ($r=1+w$) boundaries. $m$ decreases until it saturates to a value $m^* \equiv 2 \pi(1+w)/\lambda^*$.
(d) shows data collapse on a log-log scale of measured $\lambda^*$, along with $\lambda_{\rm hierarchy}$ for the pinned version (see Sec.~\ref{sec:results_C}), onto the predicted scaling law given in Eq.~\ref{eq:lambda*}. The data are for multiple numerical annuli with varying $w$, $t$ and $\Delta$. The black line represents the best-fit equation: $y = 2.2\, x$ (log-log scale). Henceforth, we term this the ``wrinklon line''.
}
\label{fig:fig1}
\end{figure*}
%%%%%%%%%%%%%%%%%%%%
%%%%%%%%%%%%%%%%%%%%
\par
In this paper, we address the question of wavelength determination in precisely such a case, for the annular geometry reported in Pal \textit{et al.}\cite{Pal2022}. The system considered there is a modification of the classic Lamé radial wrinkling deformation (see Figs.~\ref{fig:fig1}a and b), where we quasi-statically contract the inner boundary of a circular annulus through a radial displacement $\Delta$, keeping the outer boundary free. We call this the ``inner Lamé'' system. Such gradual boundary-induced contraction of the flat annulus deforms it smoothly into a pattern of uniform radial wrinkling (Fig.~\ref{fig:fig1}b), as it follows the local energy minimum.
\par
In \cite{Pal2022}, we show that this wrinkling deformation is well-modelled, even up to large amplitudes, as a (piecewise) developable surface of triangles and cones. Thus, this  deformation becomes \textit{isometric} (i.e., unstretched) as the thickness becomes much smaller than the wavelength, with its radial tension field becoming negligible as compared to that of a similarly deformed classical Lamé annulus \cite{Davidovitch2011}.
However, even in this isometric limit, where there should be no dependence on thickness, the wrinkling deformation selects a wavelength $\lambda$ that is observed to depend on both the sheet thickness $t$ and width $w$ (see Fig.~\ref{fig:fig1}d). Since the sheet is ``tension-free'', this $\lambda$ cannot be explained by the tensile mechanism of classical Lamé wrinkling \cite{Davidovitch2011}. Also, since the contracted inner boundary is both unclamped and unpinned (i.e., free to both displace and rotate out-of-plane), it cannot act to determine $\lambda$ as in references \cite{Davidovitch2009, Huang2010, Schroll2011, Vandeparre2011}.   
%that is not determined by the property of isometry. 
\par
In this paper, we show that the observed inner Lamé wavelength is consistent with a similar coarsening mechanism as in Vandeparre et al. \cite{Vandeparre2011}, but without any spatial hierarchy involved. Instead, the coarsening takes place progressively over the course of the deformation (see Fig.~\ref{fig:fig1}c), and is arrested due to the non-zero stretching energy associated with `wrinklons' -- Y-shaped spatial features where two wrinkles merge into one -- which constitute the basic unit of the coarsening process. Thus, \textit{kinetically arrested coarsening} determines the final wavenumber (i.e., the number of wrinkles) $m^*$ in the sheet. However, if the sheet manages to attain a coarser wavenumber $m < m^*$ by any means (e.g., manual setting by the experimenter), then it stays there. Thus, the wavelength selection depends on initial conditions, and can be considered \textit{hysteretic} or path-dependent. These are the central results of this paper.
\par
Below, we derive the discussed arrest criterion for the wavelength selection and demonstrate its viability numerically. The paper is organised as follows. Section \ref{sec:methods} defines the inner Lamé deformation that we simulate, and the numerical methods we use to this end. In Section \ref{sec:RESULTS}, we state the main results of this paper, deriving a scaling law for the critical arrest wavelength, $\lambda^* \sim 1/m^*$ (the black line in Fig.~\ref{fig:fig1}d), and argue that it causes hysteretic wavelength selection. Finally, Section \ref{sec:DISCUSSION} discusses the significance of these results.

%%%%%%%%%%%%%%%%%%%%
\section{Methods \label{sec:methods}}
%[Give 1. Description of annulus system, with basic parameters. Then 2. numerical methods, with 3. a separate section for boundary conditions.]
The system under consideration here is sketched in Fig.~\ref{fig:fig1}a. We start with a flat circular annulus of width $w$ and thickness $t$, whose initial inner radius we take as our unit of length. To deform this annulus, we pull the inner boundary radially inward by a distance $\Delta$, so that it is forced to live on a cylinder whose radius is reduced by $\Delta$. This leads to a contraction of the inner boundary by a factor $\Delta$, and the equilibrium shape of the sheet thus obtained is the radial wrinkling shown in Fig.~\ref{fig:fig1}b.
The system can thus be conveniently defined using only three geometric parameters: thickness $t$, width $w$, and (dimensionless) radial contraction $\Delta$. The internal forces determining the shape arise from the in-plane stretching modulus $Y = E t$, and the bending modulus $B = Et^3(1-\nu^2)/12$, where $E$ and $\nu$ are the material's Young's modulus and Poisson ratio, resp. In thin sheets with thickness $t$ much smaller than the inner radius and the width $w$, the sheet may be taken as virtually unstretchable, and many features of the shape are independent of $Y$ \cite{Davidovitch2011}. Indeed, in \cite{Pal2022}, we show that the radial wrinkling morphology under study here (see Fig.~\ref{fig:fig1}b) is isometric, such that its energy is approximately given by 
\begin{equation}\label{eq:U_sheet}
    U_{\rm sheet} \approx B m^2 \Delta.
\end{equation}
% $U_{\rm sheet} \approx B m^2 \Delta$where the approximation is for large $m$.
%\textit{geometrically}, i.e.~developably, isometric 
%(as opposed to the class of asymptotic, non-developable isometries discussed in \cite{Paulsen2019} and references therein). The macroscopic energy thus lives purely in the bending mode.

\subsection{Numerical methods}
To investigate the wrinkling morphology of the inner Lamé system here, we use the same numerical methods used in \cite{Pal2022}. For our simulations, we used the commercial finite-element (FE) solver Abaqus 2018 (SIMULIA, Dassault Systèmes). We used a protocol in which we gradually displaced the boundary in time, allowing the system to relax quasi-statically, such that the rate of motion of the boundary does not affect the emergent shape or its energy. 
The forward time integration was done using one of two standard FE protocols -- `dynamic, explicit' and `dynamic, implicit' -- both being found to give comparable results. Such dynamic integration protocols introduce inertia into the simulation, which is essential for accurately tracking local minima through instabilities and bifurcations, and thus finding reliable post-buckling solutions \cite{Taylor2015}. 
%Moreover, inertia also has a stabilising effect on the dynamics \cite{Taylor2015}. 
In contrast, static (i.e., zero inertia) simulation protocols are known to remain stuck near the initial conditions, often far away from the true minimum \cite{Taylor2015}.
However, to avoid gross kinetic effects in the presence of inertia, we made sure to increase the contraction $\Delta$ slowly enough such that the kinetic energy of the system always remained $\lessapprox 5\%$ of the elastic energy. This is standard procedure for quasi-static analyses in finite-element simulations.
The FE simulation details are given in depth in the SI of \cite{Pal2022}. Here, we give a brief overview.
\par
To simulate the inner Lamé system, we used its defining boundary conditions: radial displacement $e_r(r=1)=-\Delta$ at the inner boundary, and the outer boundary at $r=1+w$ free. We chose to apply a maximum contraction of $\Delta_{\rm max}=0.267$ at the inner boundary. All observed coarsening and selection in our simulations occurs for $\Delta$ much smaller than this maximum (see Fig.~\ref{fig:fig1}c); increasing $\Delta$ further within this range only changes the amplitude without affecting the wavenumber. We did not extend the range enough to observe the anticipated weak dependence of $\lambda^*$ on $\Delta$.
%This choice of $\Delta_{\rm max}$ was found to be amply large to accommodate all wrinkle coarsening and the selection of a stable wavenumber $m^*$ (see Fig.~\ref{fig:fig1}c); increasing $\Delta$ further only changed the amplitude without affecting the wavenumber. 
\par
To account for the possibility of high strains at such large contractions, the sheet was modelled as a Neo-Hookean hyperelastic material with coefficients equivalent to the linear moduli: Young's modulus, $E=0.907125$  MPa, and Poisson ratio, $\nu =0.475$, corresponding to a rubber-like material. To verify that results are independent of the material model, we also re-performed several simulations with a linear material model with these same moduli. 
\par
To test the validity of our results over a range of parameters, we kept the inner radius fixed and varied the other two parameters -- width $w$ and thickness $t$ -- over the range of a decade. For the width, we used values $w = 0.20, 0.33, 0.67, 1.0, 1.67$ (a factor of almost 10, ranging from very narrow to moderately wide), and for thickness, we used values $t=2.67\times10^{-3}, 1.33\times10^{-3}, 6.67\times10^{-4}, 2.67\times10^{-4}, 1.33\times10^{-4}$ (a factor of 20, ranging from moderately thick to very thin). We performed consistency checks to ensure that the final morphology was independent of the choice of any simulation parameters. 
%For data extraction, we used the software package Abaqus2Matlab \cite{Papazafeiropoulos2017}. 
%\par
%We used two principal deformation protocols to ensure that the wrinkles we observed were characteristic of the quasi-static limit. Our baseline protocol was done using explicit dynamics, in which the finite elements are accelerated with a constant damping factor. The other deformation protocol used an implicit scheme for integrating the forces, with the damping factor automatically selected by the software to favour quasi-static behaviour. In both cases, to avoid gross kinetic effects, we increased $\Delta$ from 0 to $\Delta_{\rm max}$ at a rate slow enough such that the kinetic energy of the system always remained $\lessapprox 5$\% of the elastic energy. This is standard procedure for quasi-static analyses in finite-element simulations. For explicit dynamics, we added an extra step of continuing the simulation 10\% longer, after reaching the target value of $\Delta_{\rm max}$, in order to relax any remnant kinetic energy to negligible levels.
\subsection{Boundary conditions}\label{sec:methods_B}
As noted in the Introduction, the specifics of the wrinkling morphology depend strongly on how the contracted boundary is constrained. Clamping the boundary leads to the wrinkling hierarchy morphologies discussed in \cite{Vandeparre2011} (see also Figs.~\ref{fig:fig2}b and \ref{fig:fig4}). The other extreme is to allow each point on the boundary circle to lie anywhere on the confining cylinder; such boundaries lead to wrinkling only in a transient regime, leading ultimately to collapse into a macroscopic fold. Here we study an intermediate constraint in which points on the bounding circle may displace only axially (i.e.~vertically) on the bounding cylinder; azimuthal ($\theta$) motion is not allowed. Doing so automatically prohibits folding (since this requires lateral motion), but without interfering with the wrinkling. 
%%%%%%%
\subsection{Obtaining sinusoidally-biased flat states}\label{sec:methods_C}
In Sec.~\ref{sec:results_B}, we use starting configurations with a pre-determined wavelength at the outer boundary, $\lambda_{\rm init}$. To obtain these, we biased the initial flat state with linear perturbations (of amplitude $\sim 10^{-4} - 10^{-3}$) of sinusoidal modes of known wavenumber $m$. These modes were obtained using standard linear eigenmode analysis methods in Abaqus \cite{Abaqus2018}.
%%%%%%%%%%%%%%%%%%%%
%%%%%%%%%%%%%%%%%%%%
\section{Results}\label{sec:RESULTS}
%\section{Determining $\lambda^*$ using wrinklons}
The elastic energy of the wrinkled annulus (Eq.~\ref{eq:U_sheet}) is dominated by bending. Thus, it favours the coarsest possible wrinkling (i.e., with minimum $m$). However, from Fig.~\ref{fig:fig1}d, we see that the observed wavenumber $m^*$ in the system becomes indefinitely large as we decrease thickness $t$ or width $w$. Thus, there exists some `constraint' (generically speaking) that competes with the bending energy to select the intermediate wavenumbers $m^*$. In this section, we consider this selection mechanism.
\par
%Since the wrinkles are not all uniform, the \textit{integer} variable $m^*$ is an inaccurate measure of wrinkle size. 
In what follows, instead of simply counting the number of wrinkles $m$ in the annulus, we measure the \textit{continuous} (in-plane) wavelength $\lambda$ at the outer boundary, averaged over multiple wrinkles of the sample. \footnote{It also allows us to take measurements for only half or quarter of the annulus. This makes data extraction faster.} The standard deviation in $\lambda$ then also gives a measure of the non-uniformity of wrinkling, something which the integer $m$ cannot capture. Thus, in what follows, we aim to predict the continuous variable $\lambda^*$, related to the counted $m^*$ by the approximate relationship: $\lambda^* \approx 2\pi (1+w)/m^*$.
%Instead, we use the average \textit{continuous} wavelength measured at the outer boundary: $\lambda^* \approx 2\pi (1+w)/m^*$, which is simply proportional to the inverse wavenumber. The variance in $\lambda^*$ measured over multiple wrinkles also enables us to determine the non-uniformity of the wrinkling.  
%%%%%%%%%%%%%%%%%%%%%%%%%%%%%%%%%%%%%%%%%%%%%%%%%
%%%%%%%%%%%%%%%%%%%%%%%%%%%%%%%%%%%%%%%%%%%%%%%%%
\subsection{Wrinklons and arrested coarsening}\label{sec:results_A}
\begin{figure*}[htb]
\centering
\includegraphics[width=0.85\textwidth]{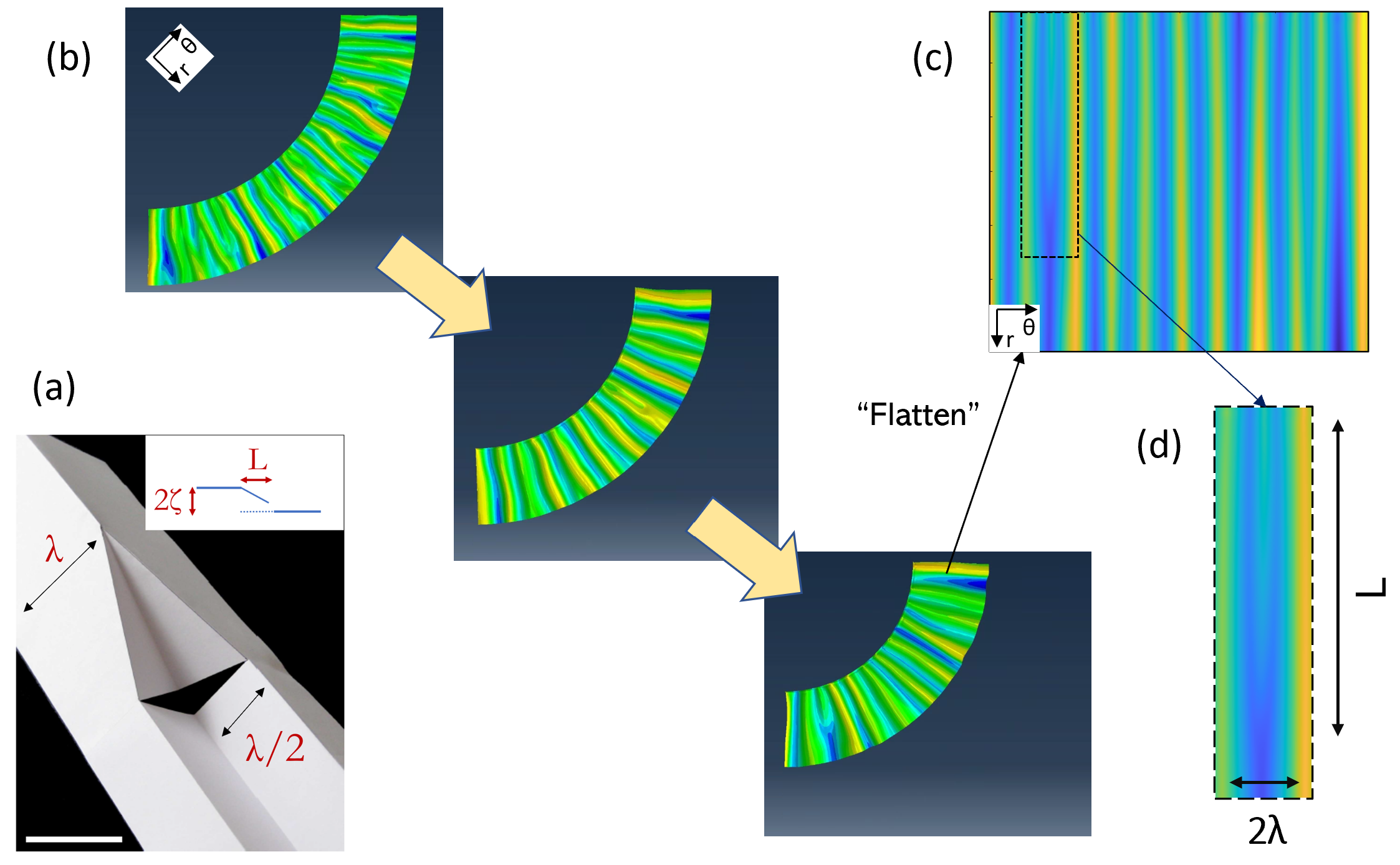}
\caption[Wavelength coarsening and wrinklons]
{Wavelength coarsening and wrinklons (see Sec.~\ref{sec:results_A}).
(a) An origami paper model of a Y-shaped wrinklon joining, wavelength $\lambda \to 2\lambda$ (measured in-plane; half-wavelengths shown), revealing the nature of its stretching (image modified from \cite{Vandeparre2011}; scale bar is 5 cm). The gap in the shape shows that a continuous sheet requires longitudinal stretching to accommodate the wrinklon (see Sec.~\ref{sec:RESULTS}B). (inset) A schematic diagram of the mid-line of the sheet (in blue), showing its vertical amplitude $\zeta$ and length $L$, used in Eq.~\ref{eq:Uwrinklon}.
(b) An illustrative example of wavelength coarsening in the inner Lamé system (for $w=0.33, t=6.67\times 10^{-4}$). The three snapshots are for the bottom-right quadrant of the annulus (coloured by height) taken at $\Delta \approx 0.07, 0.15, 0.23 $ resp.
(c) To visualise the wrinkling pattern with greater clarity, we `flatten' the polar coordinates, so that the height profile can be plotted in a rectangular matrix form (same colours, but different colour map). Here, the top edge is the inner boundary.
(d) A close-up of a wrinklon in (c), joining wavelengths $\lambda \to 2\lambda$, of longitudinal size $L$. 
% $L_{\rm wrinklon} (\lambda)$ (see Eq.~\ref{eq:Lwrinklon}). 
%Since paper is virtually unstretchable, we need a transverse cut to relieve the longitudinal extension needed to form a wrinklon
}
\label{fig:fig2}
\end{figure*}

We first note that the selection of the final wavenumber in a simulation is not an instantaneous process. Figs.~\ref{fig:fig1}c and \ref{fig:fig2}b show that, as we apply the contraction $\Delta >0$, the flat annulus initially buckles to form fine-scale radial wrinkling. This then coarsens rapidly, reducing the bending energy, but with the coarsening stopping at some final wavenumber $m^* \sim 1/\lambda^*$. This suggests that there is a continuous `path' from the initial fine value of $\lambda$ to the final value $\lambda^*$, but not beyond. Here, we argue that this path is blocked because a further coarsening becomes energetically unfavourable. In other words, the coarsening process gets kinetically arrested.
%%%%%%%%%%%%%%%%%%%%
%%%%%%%%%%%%%%%%%%%%
\par
The coarsening process is controlled by the Y-shaped spatial features named `wrinklons' \cite{Vandeparre2011}, which form every time two wrinkles merge into one. 
In what follows, we discuss the energetics of wrinklons,  approximating the wrinkles as being rectilinear (as in Fig.~\ref{fig:fig2}a), and ignoring any radial splay. We discuss the accuracy of this assumption in Sec.~\ref{sec:DISCUSSION}. 
The energetics of a wrinklon is closely associated to its shape. Fig.~\ref{fig:fig2}a shows the geometry of a wrinklon using an origami model (adapted from \cite{Vandeparre2011}). Here, the paper sheet is creased to have wavelength $\lambda$ on the right and $2 \lambda$ on the left. The transition zone is the wrinklon. As can be seen from the cut in the sheet, this involves non-zero extension of the material: to allow a trough in the $\lambda$ wave on the right to rise up to the peak of the $2\lambda$ wave on the left, a horizontal length $L$ (see inset of Fig.~\ref{fig:fig2}d) must be stretched into the hypotenuse $\sqrt{(2\zeta)^2 +L^2}$, where $\zeta$ is the amplitude of the $\lambda$-wave. For small slope $\zeta/L$, this generates a strain of order $(\zeta/L)^2$. In the absence of a substrate or boundary tension, this contributes a stretching energy density $\sim Y (\zeta/L)^4$, where $Y$ is the stretching modulus. 
%Simultaneously, the wrinklon has a bending energy density $\sim B (\zeta/\lambda^2)^2$, where $B$ is the bending modulus. 
\par
Since the inner Lamé wrinkling is isometric or strain-free \cite{Pal2022}, we can use the constraint of inextensibility (i.e., length conservation) 
%Furthermore, since the material circles are unstretched in the wrinkled configuration, one can use  
to relate the average slope $\zeta/\lambda$ of the wrinkled circles to the applied contraction $\Delta$. Taking this latter to be constant over the length of the wrinklon, for small slopes, we have: $(\zeta/\lambda)^2 \sim \Delta$. This is popularly known in literature as the `slaving condition' \cite{Paulsen2019}, since it shows that the amplitude $\zeta$ and wavelength $\lambda$ are co-dependent variables for inextensible wrinkling. Thus, removing $\zeta$ in favour of $\lambda$ and $\Delta$, the elastic energy of a wrinklon of area $\sim L \lambda$ is given by:
\begin{equation}\label{eq:Uwrinklon}
   U_{\rm wrinklon} \sim Y \lambda^5 L^{-3} \Delta^2.
\end{equation}
% \begin{align}\label{eq:Uwrinklon}
%     U_{\rm wrinklon} &\sim Y \lambda^5 L^{-3} \Delta^2 + B L \lambda^{-1} \Delta \\ \nonumber
%     &\sim Y (\lambda^5 L^{-3} \Delta^2 + t^2 L \lambda^{-1} \Delta),
% \end{align}
% where we have used $B \sim Y t^2$.
% In Eq.~\ref{eq:Uwrinklon}, we see that the stretching and bending energies scale with length scale $L$ in opposite ways. While the bending energy contribution decreases at smaller $L$, the stretching energy \textit{sharply} increases. Thus, at the equilibrium length scale
% \begin{equation}\label{eq:Lwrinklon}
%     L_{\rm wrinklon}(\lambda) \sim \lambda^{3/2} t^{-1/2} \Delta^{1/4},
% \end{equation}
% the stretching and bending energies balance each other, and minimise $U_{\rm wrinklon}$. Eq.~\ref{eq:Lwrinklon} identifies the characteristic longitudinal length scale associated with a wrinklon of transverse size $\lambda$. We note that similar examples of stretching localised to characteristic length scales (known as ``stress concentration'' \cite{Witten2007}) can be seen in other systems \cite{Lobkovsky1997, Cerda1998, Cerda1999, Mahadevan2007}. 
\par
To see how such wrinklons can select the wavelength, consider the following picture of kinetic arrest.
% The length-based argument for wavelength selection above can be reinforced by an energetic picture. 
Assume uniform wrinkling of wavelength $\lambda$ over the entire width $w$ of the sheet. Then this wavelength can coarsen further if and only if a wrinklon can form. This will happen if and only if there is a decrease in the net elastic energy during this process. Since there is an energy gain from creating the wrinklon, as well as a bending energy decrease due to wrinkle coarsening from $\lambda \to 2\lambda$, the net change in energy $\delta U$ is:
\begin{equation}\label{eq:deltaU_definition}
    \delta U (\lambda) = U_{\rm wrinklon} (\lambda) - \delta U_{\rm bend} (\lambda),
\end{equation}
where $\delta U_{\rm bend} (\lambda) \sim  B \Delta w \lambda^{-1}$ for an area $\sim \lambda w$. Thus, a wrinklon will form, 
%move to the inner boundary (see Fig.~\ref{fig:fig2}b), 
and the wavelength will increase, only if $\delta U (\lambda) < 0$. 
Setting $\delta U (\lambda) = 0$ then gives us a critical minimum scale $L^* (\lambda)$ (using Eq.~\ref{eq:Uwrinklon}):
\begin{equation}\label{eq:Lstar_scaling}
    L^* (\lambda) \sim t^{-2/3} \Delta ^{1/3} w^{-1/3} \lambda^2,
\end{equation}
such that only wrinklons having length $L > L^*(\lambda)$ are energetically feasible in the system, i.e., they lower the net energy: $\delta U (\lambda) < 0$. 
Since the wrinklon has to be smaller than the sheet's width as well, we find that wrinklons can form as long as their size $L$ obeys the bounds: $w > L > L^*(\lambda)$.
\par
However, with each round of coarsening, as $\lambda$ increases, Eq.~\ref{eq:Lstar_scaling} shows us that $L^* (\lambda)$ also increases (rapidly). Thus, when
$L^*(\lambda) > w$, there can be no wrinklon that reduces the energy, so that no further coarsening can occur. This defines a critical wavelength scale $\lambda^*$ such that
\begin{equation}\label{eq:Lstar_critical}
    L^*(\lambda)|_{\lambda^*} = w.
\end{equation}
Using Eq.~\ref{eq:Lstar_scaling}, we get:
\begin{equation}\label{eq:lambda*}
     \lambda^* \sim w^{2/3} t^{1/3} \Delta^{-1/6}.
\end{equation}
% Using Eqs.~\ref{eq:Uwrinklon} and \ref{eq:Lwrinklon} for $U_{\rm wrinklon} (\lambda)$ above, we find that this energetic inequality translates to a length inequality:
% \begin{equation}\label{eq:Lwrinklon_inequality}
%     L_{\rm wrinklon}(\lambda) \lesssim w,
% \end{equation}
% with the arrest criterion being given by
% \begin{equation}\label{eq:arrestCriterion_length}
%     L_{\rm wrinklon}(\lambda) |_{\lambda^*} \sim w.
% \end{equation}
% The critical wavelength scale $\lambda = \lambda^*$ is obtained by inverting Eq.~\ref{eq:arrestCriterion_length} using Eq.~\ref{eq:Lwrinklon}: 
% \begin{equation}\label{eq:lambda*}
%      \lambda^* \sim w^{2/3} t^{1/3} \Delta^{-1/6}.
% \end{equation}
The conclusion is that all wavelengths $\lambda$ with $L^*(\lambda) < w$, and thus $\lambda \lesssim \lambda^*$, are susceptible to wrinklon-mediated coarsening, and are hence unstable and \textit{should not be visible} in the sheet. Moreover, these finer wavelengths should get coarsened up to or just above the critical wavelength $\lambda^*$; beyond this, coarsening is energetically unfeasible and gets arrested. 
This is confirmed in Fig.~\ref{fig:fig1}d, where the best-fit $\lambda^*$ line is drawn in black (henceforth, we call this the `wrinklon line').
\par
Eq.~\ref{eq:lambda*} is the central prediction of this paper. It emerges directly from Eq.~\ref{eq:Lstar_critical}, which represents an arrest criterion for wrinklon-mediated coarsening in the annulus. This shows that \textit{transient} wrinklons in the inner Lamé system can select a wavelength by a mechanism of kinetic arrest.
This wavelength $\lambda^*$ is an emergent phenomenon depending on all three geometric factors $t$, $w$ and $\Delta$, but independent of material constants $E$ and $\nu$. 
The pre-factor is a universal number (estimated to be $\approx 2.2$), independent of initial wavelength. This is much as in \cite{Vandeparre2011}, where the outer wavelength determined through spatial coarsening is independent of the value at the clamped inner boundary.
\par
Eq.~\ref{eq:lambda*} predicts wider and thicker sheets to display coarser wrinkling, and vice versa for narrower, thinner sheets.
On the other hand, the arrest argument above also suggests that wavelengths coarser than $\lambda^*$ should remain stable since they are not subject to further coarsening via wrinklons. Our next step is thus to confirm the possibility of the sheet supporting wavelengths $\lambda \gtrsim \lambda^*$.
%%%%%%%%%%%%%%%%%%%%%%%%%%%%%%%%%%%%%%%%%%%%%%%%%
%%%%%%%%%%%%%%%%%%%%%%%%%%%%%%%%%%%%%%%%%%%%%%%%%
\subsection{Attaining coarser wavelengths by changing initial wavelength}\label{sec:results_B}
The natural tendency of the wrinkled annulus is to minimise its bending energy and coarsen as much as possible. The arrest argument above suggests that, \textit{using wrinklons}, the wrinkling cannot coarsen beyond $\lambda^*$. However, given an opportunity to bypass the wrinklon mechanism and coarsen directly, the annulus should do so. 
%%%%%%%%%%%%%%%%%%%%
%%%%%%%%%%%%%%%%%%%%
\begin{figure*}[htb]
\centering
\includegraphics[width=0.9\textwidth]{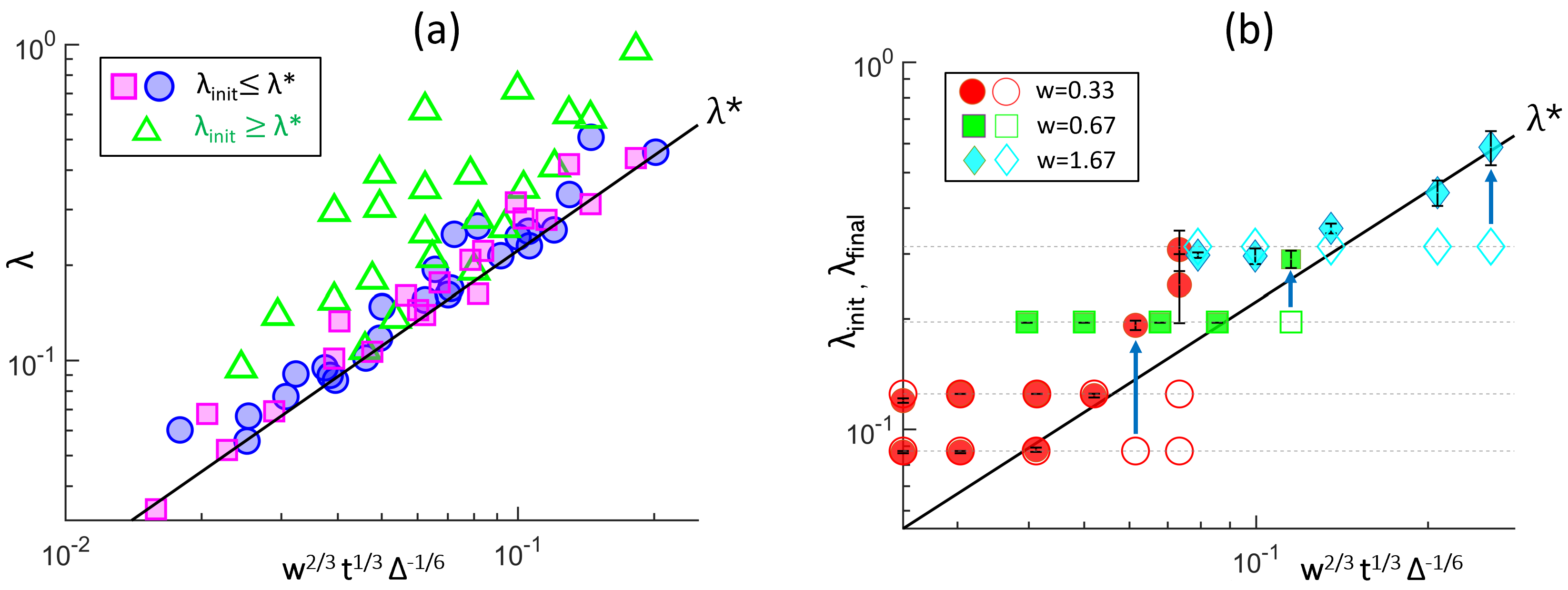}
\caption[Effect of varying the initial wavelength]
{Effect of varying the initial wavelength $\lambda_{\rm init}$. (a) The data points (in blue and magenta) and best-fit `wrinklon line' from Fig.~\ref{fig:fig1}c, having $\lambda_{\rm init} \leq \lambda^*$, are overlaid with new sample points (in green) that start from a \textit{biased} flat state with a perturbation of wavelength $\lambda_{\rm init} \geq \lambda^*$. These biased samples show little to no coarsening. (b) We see the difference in coarsening behaviour clearly by plotting both $\lambda_{\rm init}$ (empty symbols) and the final wavelength $\lambda_{\rm final}$ (solid symbols) for some selected samples. Here, we fix $\lambda_{\rm init}$ and width $w$, and vary the thickness $t$, thereby creating a horizontal row of empty symbols for given $\lambda_{\rm init}$. We plot $\lambda_{\rm init}$ and $\lambda_{\rm final}$ with the same $w$, $t$ and $\Delta$, so that any coarsening is noticeable by a vertically upward shift of the solid symbol (some arrows drawn for indication). We see a clear transition from non-coarsening to coarsening behaviour as we cross the black wrinklon line. The error bars denote standard error from averaging over all the wrinkles.
}
\label{fig:fig3}
\end{figure*}
%%%%%%%%%%%%%%%%%%%% 
%%%%%%%%%%%%%%%%%%%%
\par
One way to test this hypothesis is to \textit{manually} set a coarser wavelength in our simulations. We do this by biasing the initial flat state (see Sec.~\ref{sec:methods_C}) with a sinusoidal perturbation of known outer wavelength $\lambda_{\rm init}$. In Fig.~\ref{fig:fig3}a, we show the results of this method. The blue and magenta dots are the same data points from Fig.~\ref{fig:fig1}c, showing the final wavelength $\lambda$ for annuli contracted from a flat or nearly flat state (i.e., having $\lambda_{\rm init} \ll \lambda^*$) or a state biased with sinusoidal modes having $\lambda_{\rm init} \lesssim \lambda^*$. They all lie on the black `wrinklon line': $\lambda = \lambda^*$. On the other hand, the new green dots are obtained by starting from biased states having $\lambda_{\rm init} \gtrsim \lambda^*$. We find that, on average, these samples do not coarsen at all. In other words, they are not susceptible to the wrinklon coarsening mechanism. Indeed, we used this method in \cite{Pal2022} to generate uniformly wrinkled patterns to provide clean geometric data. Thus, Fig.~\ref{fig:fig3}a is fully consistent with the hysteretic picture posited at the end of Sec.~\ref{sec:results_A} 
\par
In Fig.~\ref{fig:fig3}b, we present specific examples of this $\lambda$-dependent coarsening. Here, we compare the initial ($\lambda_{\rm init}$) and final ($\lambda_{\rm final}$) wavelengths of samples, as we vary their parameters. Specifically, for fixed initial wavelength $\lambda_{\rm init}$ (i.e., the ordinate), we change the abscissa by varying the thickness $t$ for samples of fixed width $w$ and contraction $\Delta$, and record the simulation $\lambda_{\rm final}$. 
Fig.~\ref{fig:fig3}b shows some representative data points, showing both $\lambda_{\rm init}$ (empty symbols) and $\lambda_{\rm final}$ (full symbols) for these samples. We see that the leftmost (i.e.~thinnest) samples, which start above the wrinklon line, do not coarsen. However, as soon as we cross the wrinklon line horizontally (following the dotted grey line), the samples start to coarsen, i.e., they move straight up. As expected, the coarsening happens up to or above the wrinklon line, consistent with the discussion in Sec.~\ref{sec:results_A}. 
\subsection{$\lambda^*$ starting from wrinkle hierarchy}\label{sec:results_C}
%%%%%%%%%%%%%%%%%%%%
\begin{figure*}[htb]
\centering
\includegraphics[width=0.8\textwidth]{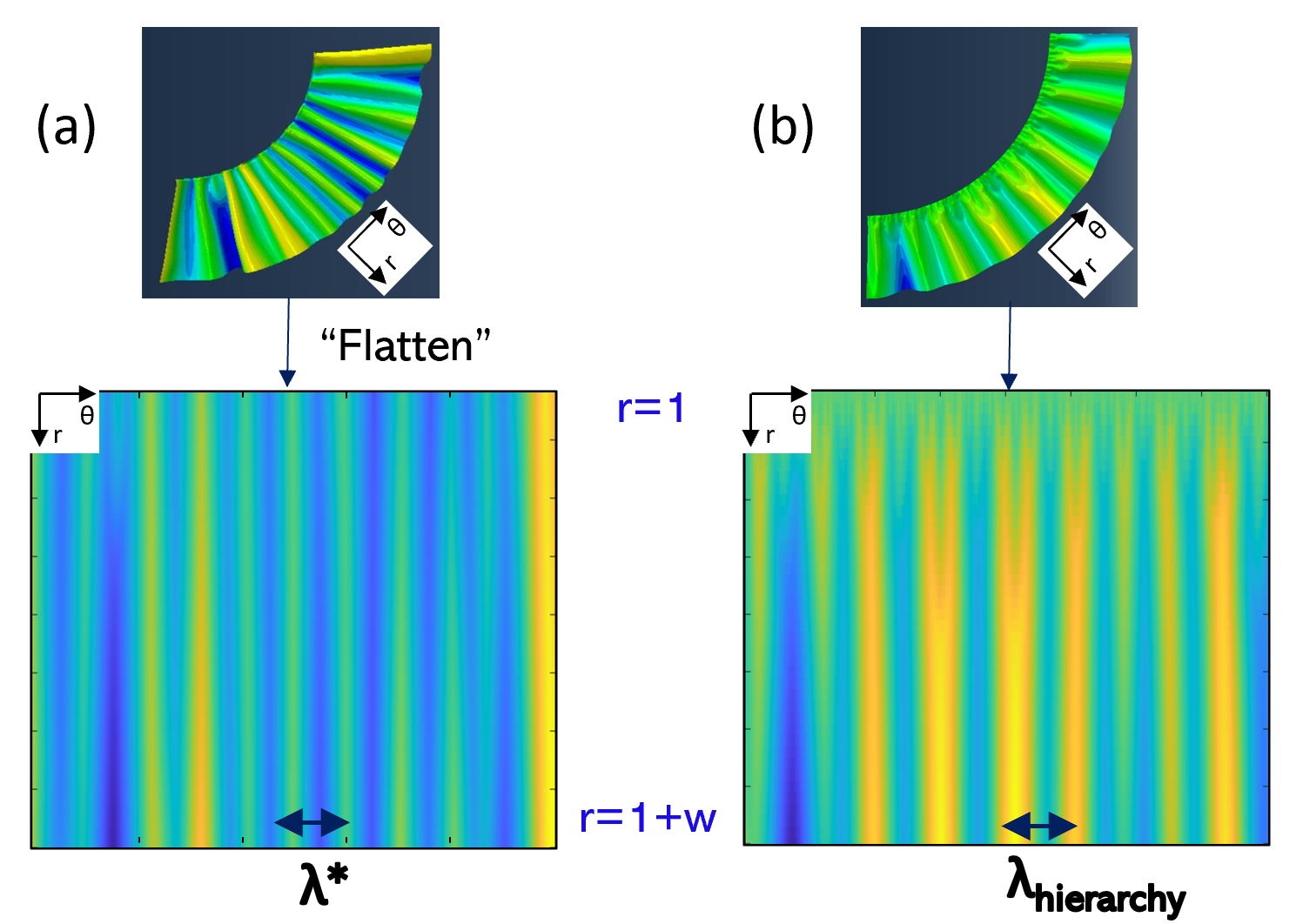}
\caption[Contrasting wrinkled morphologies but similar outer wavelengths]
{Contrasting wrinkled morphologies but similar outer wavelengths, for two different boundary conditions at the inner boundary ($r=1$) (see Sec.~\ref{sec:results_C}). Data are for the same representative sample used in Fig.~\ref{fig:fig2}b ($w=0.33$, $t=6.67\times10^{-4}$). The top row shows the bottom-right quadrant of the deformed annulus in both cases (coloured by height). The bottom row shows the same height profile, but flattened into a rectangular matrix form, as in Fig.~\ref{fig:fig2} above.
(a) The ``inner Lamé'' boundary condition, the subject of this paper, which allows free vertical displacement at $r=1$. This leads to near-uniform radial wrinkling, with wavelength $\lambda^*$ at the outer boundary ($r=1+w$). (b) A pinned boundary condition that prohibits vertical displacement at $r=1$. This leads to a spatial wrinkle hierarchy, that terminates with the coarsest wavelength $\lambda_{\rm hierarchy}$ at $r=1+w$. 
}
\label{fig:fig4}
\end{figure*}
%%%%%%%%%%%%%%%%%%%%
The inner Lamé deformation (Fig.~\ref{fig:fig1}b) emerges from contracting an annulus whose inner boundary is free to displace vertically. The arrest-based mechanism described in the previous two sub-sections suffices to explain wavelength selection in this system. To better understand this mechanism, however, it is instructive to consider a different system: one where the inner boundary ($r=1$) is first \textit{pinned} -- thereby prohibiting any vertical displacement -- then contracted by the same fraction $\Delta$, and then unpinned again. Thus, the final state is that of the inner Lamé system -- namely, having an unpinned inner boundary contracted by $\Delta$ -- but the path taken to arrive there is different.
\par
The reason this parallel, pinned system is instructive is because it allows us to make a direct connection to a class of known systems. 
%While the previous two sub-sections suffice to describe the central results of this paper, it is instructive to see the selection of $\lambda^*$ from a different perspective, as emerging smoothly from the modification of a more constrained version of the inner Lamé system. In this version, we pin the inner boundary ($r=1$), prohibiting any vertical, out-of-plane displacement. 
As shown in Fig.~\ref{fig:fig4}b, applying contraction $\Delta$ with pinned boundary conditions (BCs) leads to the creation of a spatial wrinkle-hierarchy morphology -- like the ones studied in \cite{Huang2010,Vandeparre2011, Schroll2011,Davidovitch2009} for rectangular geometries – where fine-scale wrinkling at the pinned boundary coarsens progressively in space via multiple generations of wrinklons. Despite the qualitatively different morphologies between the pinned and unpinned cases (see Fig.~\ref{fig:fig4}), we argue that one should still expect the wavelengths at the outer boundary for the two cases to be the same. To see this, consider the following.
%-- $\lambda^*$ for the unpinned case and $\lambda_{\rm pinned}$ for the pinned one --
\par
We can think of the emergence of the unpinned morphology in Fig.~\ref{fig:fig4}a starting from the pinned, wrinkle-hierarchy case in Fig.~\ref{fig:fig4}b. Eq.~\ref{eq:Uwrinklon} tells us that wrinklons cost non-zero elastic energy. Moreover, they separate a region of fine, higher-energy wrinkling (nearer $r=1$), from a region of coarser, lower-energy wrinkling (nearer $r=1+w$). Thus, allowing the wrinklons to move right up to the inner boundary would eliminate not only the wrinklons themselves but also the entire fine wrinkling region, leading to a net lowering of the sheet's energy.

%Thus, eliminating wrinklons from a configuration should lead to a lowering of its energy. 
Imagine gradually undoing the pinning at $r=1$ in Fig.~\ref{fig:fig4}b, allowing the boundary nodes to displace vertically within a maximum height $\epsilon$. As $\epsilon$ increases, the boundary will be able to support coarser wavelengths.
Thus, as it reaches the approximate height of the innermost generation of wrinklons, in order to minimise elastic energy, these wrinklons should all migrate to the boundary and disappear, thereby increasing the wavelength at $r=1$ by a generation. Similarly, when $\epsilon$ reaches the height of the second (now innermost) wrinklon generation, this generation should also vanish in a similar manner, decreasing the energy and coarsening the wavelength by a further generation. As $\epsilon$ increases further, this coarsening process may continue as long as there are wrinklons that can move into the inner boundary. That is, as long as there are more inner wrinkles than outer ones. The end point of this process is thus a state where the original outer wrinkles extend to the inner boundary.
Thus, we would reach the unpinned boundary conditions of the original inner Lamé system, and anticipate the same morphology (i.e.~Fig.~\ref{fig:fig4}a). This argument tells us that we should expect:
\begin{equation}\label{eq:lambda*_from_hierarchy}
    \lambda^* = \lambda_{\rm hierarchy} (x=w),
\end{equation}
where $\lambda_{\rm hierarchy} (x)$ is the wavelength at distance $x$ from the pinned boundary in the wrinkle hierarchy. Using the expression 
\begin{equation}\label{eq:lambda(x)_vandeparre}
    \lambda_{\rm hierarchy} (x) \sim x^{2/3} t^{1/3} \Delta^{-1/6}
\end{equation}
given for rectangular geometries in Vandeparre et al.~\cite{Vandeparre2011}, we recover Eq.~\ref{eq:lambda*} for $\lambda^*$. To verify the prediction of Eq.~\ref{eq:lambda*_from_hierarchy}, we perform simulations of this pinned BC for our annular systems to measure $\lambda_{\rm hierarchy} (x=w)$. We then compare $\lambda_{\rm hierarchy} (x=w)$ to $\lambda^*$ measured for the unpinned case. We find that our measurements are fully consistent with Eq.~\ref{eq:lambda*_from_hierarchy}. We can already see this qualitatively in the bottom row of Fig.~\ref{fig:fig4}. Fig.~\ref{fig:fig1}d shows this quantitatively: $\lambda^*$ and $\lambda_{\rm hierarchy}(x=w)$ measured for a wide range of samples with varying $w$, $t$ and $\Delta$, collapse on to the same straight line given by Eqs.~\ref{eq:lambda*} and \ref{eq:lambda(x)_vandeparre}. The ``unpinned-hierarchy" argument for $\lambda^*$ described in this section thereby provides an independent confirmation of the arrest criterion argument presented in Secs.~\ref{sec:results_A} and \ref{sec:results_B}.
%%%%%%%%%%%%%%%%%%%%%%%%%%%%%%%%%%%%%%%%%%%%%%%%%
%%%%%%%%%%%%%%%%%%%%%%%%%%%%%%%%%%%%%%%%%%%%%%%%%
\subsection{Summary}
The results presented in Sec.~\ref{sec:RESULTS} together paint the following story. First, the radially-contracted inner Lamé system, 
despite having its energy dominated by only bending,
%despite having only bending energy at the macroscopic scale, 
selects a critical wavelength scale $\lambda^*$ much finer than the system size. This $\lambda^*$ is determined through competition between stretching and bending energies at the scale of local features called wrinklons. Second, the selected $\lambda^*$ is a function of the single collective variable $ w^{2/3} t^{1/3} \Delta^{-1/6} \equiv \tilde x$, sensitive to both the smallest (thickness $t$) and largest (width $w$) dimensions of the sheet. Third, the wavelength selection is hysteretic (i.e.~path-dependent), since the final wavelength $\lambda$ selected by the annulus depends on its initial location in the $\lambda-\tilde x$ plane. As seen in Fig.~\ref{fig:fig3}a, the critical wavelength $\lambda^*$ acts as a linear discriminant in this plane, separating it into two regions. Consistent with its desire to minimise bending energy, the sheet can support wavelengths coarser than $\lambda^*$ but nothing finer than it. Thus, the wavelengths supported by the sheet obey the inequality: $\lambda \gtrsim \lambda^*$. 
%-- which we estimate to obey $\log \lambda^* \approx 2.2 \, \log \tilde x$ through fitting -- 
%%%%%%%%%%%%%%%%%%%%%%%%%%%%%%%%%%%%%%%%%%%%%%%%%
%%%%%%%%%%%%%%%%%%%%%%%%%%%%%%%%%%%%%%%%%%%%%%%%%
\section{Discussion}\label{sec:DISCUSSION}
\subsection{Significance}
The mechanism of \textit{wrinklon-mediated arrested coarsening} presented in this paper accounts for all the qualitative and quantitative features of the wavelengths $\lambda$ measured in our inner Lamé  samples. The novelty of this mechanism is closely related to the nature of the inner Lamé deformation. This involves quasi-static contraction of the annulus, so that it follows a local energy minimum. The deformation does not thus represent a ground state (i.e., a global energy minimum) \cite{Pal2022}, and this is reflected in the proposed wavelength selection mechanism.   
The hysteretic wavelength selection paradigm presented here is thus very different from the traditional paradigms of wavelength selection in elastic wrinkling, which are based on global energy minimisation and hence are path-independent. For example, consider the prototypical case of wrinkling in a thin sheet of bending stiffness $B$ attached to a substrate of stiffness $K$ \cite{Cerda2003}\footnote{In terms of the quantities already introduced in this paper, $K$ has units of $Y/(\rm length)^2$ or $B/(\rm length)^4$.}. Competition between the bending of the sheet and the deformation of the substrate leads to the selection of a wavelength $\lambda_{\rm substrate} \sim (B/K)^{1/4}$ that minimises the global energy. $\lambda_{\rm substrate}$ here does not depend on the initial value of the wavelength chosen. If we start with a putative wavelength $\lambda' \lesssim \lambda_{\rm substrate}$, then bending energy will force the wavelength to increase to $\lambda_{\rm substrate}$. Conversely, if we start with some $\lambda' \gtrsim \lambda_{\rm substrate}$, then the substrate energy will force the wavelength to decrease to $\lambda_{\rm substrate}$. Thus, on a plot of $\lambda$ vs.~$(B/K)^{1/4}$, data points measured from samples with different values $B$ and $K$ should all lie on a single straight \textit{line}.  
\par
In contrast, in the inner Lamé system, the lack of a macroscopic force to compete with the sheet's bending energy means that the ground state is an $m \to 0$ fold-like solution \cite{Pocivavsek2008, Brau2013}. Instead, the inner Lamé contraction
%In contrast, in the inner Lamé system, the lack of a macroscopic force to compete with the sheet's bending energy means that there is no non-zero global energy minimum.
selects a wrinkled configuration of non-zero wavenumber, which we claim is due to kinetic considerations.
This wavenumber is selected through \textit{local} competition at scale $L^*$ between bending and \textit{stretching}, this latter being the only source of competition possible in an unsupported sheet. The observed one-sided hysteresis is a direct consequence of this. Bending energy wants to coarsen wavelengths as much as possible, but the size of the sheet $w$ acts as a fundamental barrier to this coarsening, through the ``coarsening condition'': $L^* (\lambda) \leq w$. Conversely, the lack of a (real or effective) substrate means that there are no penalties to wavelengths coarser than $\lambda^*$. Thus, on the $\lambda-\tilde x$ wavelength plane, there is an entire \textit{range} of wavelengths available to the sheet: $\lambda \gtrsim \lambda^*$.
%The selection of the hysteretic discriminant $\lambda^*$ is an emergent phenomenon, sensitive to both the smallest (thickness $t$) and largest (width $w$) dimensions of the sheet.
\par
Thus, in a fundamental but non-trivial way, the wavelength $\lambda$ in the inner Lamé system is selected by the sheet's size $w$. This is similar and yet dissimilar to known cases of geometric wavelength selection in isometrically buckled systems, where \textit{both} macroscopic stretching and bending are absent. A good example is the faceted twisted ribbon in \cite{PhamDinh2016}, which buckles into a periodic pattern of triangular facets. Here, the wavelength is directly set by the ribbon's width $w$: $\lambda = w$ \cite{PhamDinh2016}. Our hysteretic wavelength selection also involves the sheet width, but through an \textit{inequality}. In this sense, it is a weaker and less restrictive geometric selection principle.
\subsection{Generalisation}
An expected consequence of such a simple \textit{size-based} selection principle (c.f.~Eq.~\ref{eq:Lstar_critical}) is that it should generalise to other cases as well, provided wrinklons are the dominant wavelength-selection mechanism. 
For example, consider wavelength selection in a hanging rectangular curtain. Let the curtain be short and light, so that gravity is negligible. Let its height be $h$, and its manually fixed wavelength at the top boundary be $\lambda_0$. If $\lambda_0$ is coarse enough that $L^*(\lambda_0) > h$, then there should be no wrinklons in the curtain, and we should see a uniform wrinkling pattern with wavelength $\lambda_0$. However, if we decrease $\lambda_0$ gradually, then we should find wrinklons forming as soon as $L^*(\lambda_0) \approx h$, with a consequent doubling of the wavelength at the outer boundary. Such a concrete prediction for a crossover can be easily verified through experiments. Given Eq.~\ref{eq:Lstar_critical}, we would expect this result even to extend to heavy curtains, but with a different scaling law for $L^*$ that includes gravity \cite{Vandeparre2011}.
%A curtain is not only rectangular, but also subject to a vertical tension field in the form of gravity. Due to this weight-induced tension, we would expect $L^*$ in curtains to scale differently \cite{Vandeparre2011} from our tension-free case (see Eq.~\ref{eq:Lstar_scaling}). Irrespective of the scaling however, \cite{Vandeparre2011} shows that the outer wavelength in a hanging curtain is selected by wrinklons. Then our size-based wavelength selection principle has the following implication. 
\par
We also note that the scaling law Eq.~\ref{eq:lambda*} for $\lambda^*$ accounts for all the parameters in our inner Lamé system. Thus the ratio of $\lambda^*$ to $\tilde x$ is a pure number ($\approx 2.2$) that should be the same even for more general cases of unsupported, inner boundary-contracted annuli (with free outer boundary). In Sec.~\ref{sec:RESULTS}, we have shown Eq.~\ref{eq:lambda*} to hold for both unpinned and pinned inner boundaries, with purely radial displacement. In Sec.~\ref{sec:methods_B}, we mention the case where the contracted boundary is also allowed to displace azimuthally on the constraining cylinder, and shows wrinkling as a transient state prior to folding \cite{Pal2022}. Here, the wrinkles persist on a time scale required for boundary points to migrate azimuthally over a finite fraction of the circumference. We would expect Eq.~\ref{eq:lambda*} to be valid in this transient regime.
Finally, we might expect Eq.~\ref{eq:lambda*} to hold even for cases where the inner boundary is contracted in a different manner, e.g., when it is forced to live on a constraining cone instead of a cylinder. However, the numerical pre-factor need not be the same in this case.
\subsection{Remarks}
Certain features of our results in Sec.~\ref{sec:RESULTS} deserve comment.
First, we note that both derivations of the threshold $\lambda^*$ (viz., Eqs.~\ref{eq:lambda*} and \ref{eq:lambda*_from_hierarchy}) ignore any splay in the wrinkle structure arising from the overall annular geometry of the system. The fact that our measurements coincide with the prediction thus suggests that wrinkle splay is unimportant in the regime of $w$, $t$ and $\Delta$ considered here. One might expect such splay to become important particularly for large width $w$.
% (Note to self: There are two possible sources for the dispersion: one is from the jump in m to m-1, which affects the mean value of $\lambda$. The other is from the variability of \lambda within a sample (i.e., non-uniform wrinkles, see figures), which generates an error bar on the mean \lambda. The two are independent effects. m to m-1 is always true. The effect is stronger for smaller m, larger \lambda (don't mention). The second explanation is due to the random nature of a coarsening evening: the fact that coarsening is an equal probability event for each wrinkle pair makes coarsening more probable if I have larger m to start out with. Here, the effect is stronger for larger m. That is why I include only the second reason.But perhaps I should include the first as well. )
Second, in Figs.~\ref{fig:fig3}a and b, we note that the data points which have coarsened up to the wrinklon line $\lambda^*$ from below show considerable dispersion. The data points reflect the mean outer $\lambda$, averaged over all wrinkles in an annulus; in Fig.~\ref{fig:fig3}b, we add error bars to additionally depict the imprecision in $\lambda$ for a sample. 
Such variability can be explained partially by two factors. The first is that the real simulation wavenumber $m$ must be an integer, and thus any wrinklon that migrates to the inner boundary must change $m$ by $-1$. This will correspond to jumps, and hence dispersion, in the measured outer wavelength $\lambda \sim 1/m$. The second factor to consider is the random nature of the coarsening events. All wrinklons do not coarsen simultaneously (thereby changing $m \to m/2$ directly). 
Instead, wrinklons form randomly and wrinkle-pair-by-pair, with wrinkle pairs separated in space, \textit{a priori}, coarsening independently. This leads to $m$ decreasing in smaller steps (see Fig.~\ref{fig:fig1}c), and also to heterogeneity in the wrinkles and wrinklons (see Fig.~\ref{fig:fig4}b and the error bars in Fig.~\ref{fig:fig3}b).
The energetic inequality $\delta U <0$ (see Eq.~\ref{eq:deltaU_definition}) only marks the feasibility regime for such coarsening events to occur. 
Consequently, if coarsening is an equal probability event for each wrinkle pair, then one should expect a larger drop in $m$, i.e.~more coarsening, if the configuration starts out with a larger initial $m$. We see that, among the coarsened data points in Fig.~\ref{fig:fig3}b, the red dots ($w=0.33$) show noticeably large coarsening. This discrepancy might then be related to the fact that these samples started from biased states with a relatively large initial wavenumber $m_{\rm init} \approx 25 - 35$, compared to $m_{\rm init} \approx 15-20$ for the wider samples (green and teal symbols).
Furthermore, we note that the sign of the discrepancy in Fig.~\ref{fig:fig3}b is generally such that the data points lie \textit{above} $\lambda^*$. This is consistent with our claim that $\lambda^*$ only acts as a \textit{lower threshold} to the wavelengths observable in the system: $\lambda \gtrsim \lambda^*$.
\par
%Finally, we note that the presence of expensive stretching, even transient, in an otherwise pure-bending configuration might seem paradoxical. 
Finally, we recall that \cite{Pal2022} shows the inner Lamé wrinkle configuration to be isometric (see Eq.~\ref{eq:U_sheet}). On the other hand, in this paper, we use stretching energy to explain the wavelength selection. This might seem paradoxical, but is so only at first glance. The stretching energy of the wrinklon defines an energy barrier that would need to be crossed in order to increase the wavelength beyond $\lambda^*$. It does not influence the energy of the state for a given $\lambda$ (equivalently, a given $m$), which is the subject of \cite{Pal2022}.
%
% This might seem paradoxical at first glance.  However, the presence of stretching in a pure-bending configuration is allowed, precisely since it is localised to the emergent length scale $L_{\rm wrinklon}$ set by balancing stretching and bending energies. This is exactly analogous to the presence of localised (viscous) boundary layers in inviscid fluids \cite{Batchelor2000}.

%%%%%%%%%%%%%%%%%%%%%%%%%%%%%%%%%%%%%%%%%
%%%%%%%%%%%%%%%%%%%%%%%%%%%%%%%%%%%%%%%%%
%%%%%%%%%% END OF MAIN TEXT %%%%%%%%%%%%%
%%%%%%%%%%%%%%%%%%%%%%%%%%%%%%%%%%%%%%%%%

%\section{Conclusions}
%The conclusions section should come in this section at the end of the article, before the Conflicts of interest statement.
%\section*{Author Contributions}
%We strongly encourage authors to include author contributions and recommend using \href{https://casrai.org/credit/}{CRediT} for standardised contribution descriptions. Please refer to our general \href{https://www.rsc.org/journals-books-databases/journal-authors-reviewers/author-responsibilities/}{author guidelines} for more information about authorship.

\section*{Conflicts of interest}
There are no conflicts to declare.

\section*{Acknowledgements}
The author would like to thank Benny Davidovitch and Enrique Cerda for many insightful discussions, and Luka Pocivavsek, Nhung Nguyen and George Papazafeiropoulos for help with numerical analysis. This work was done in pursuit of a PhD under the supervision of Prof.~Thomas A.~Witten. This work was primarily supported by the University of Chicago Materials Research Science and Engineering Center (MRSEC), funded by National Science Foundation (NSF) grants DMR-1420709 and DMR-2011854, the latter in the form of a UChicago-MRSEC student fellowship.

%%%END OF MAIN TEXT%%%


\balance


%%%REFERENCES%%%
\bibliography{paperBib_mDetermination_Lame2022} %You need to replace "rsc" on this line with the name of your .bib file
\bibliographystyle{rsc} %the RSC's .bst file

\end{document}